\documentclass[rmp,superscriptaddress,twocolumn,floatfix]{revtex4}

\usepackage{dcolumn,graphicx,amsmath,amssymb,txfonts}

\makeindex

\begin{document}

\title{The diplomat's dilemma: Maximal power for minimal effort in
  social networks}
\author{Petter Holme}
\affiliation{Royal Institute of Technology, 10044 Stockholm, Sweden}
\author{Gourab Ghoshal}
\affiliation{Department of Physics, University of Michigan,
Ann Arbor, MI 48109, U.S.A.}

\begin{abstract}Closeness is a global measure of centrality in networks,
  and a proxy for how influential actors are in social networks. In
  most network models, and many empirical networks, closeness is
  strongly correlated with degree. However, in social networks there
  is a cost of maintaining social ties. This leads to a situation
  (that can occur in the professional social networks of executives,
  lobbyists, diplomats and so on) where agents have the conflicting
  objectives of aiming for centrality while simultaneously keeping the
  degree low. We investigate this situation in an adaptive
  network-evolution model where agents optimize their positions in the
  network following individual strategies, and using only local
  information. The strategies are also optimized, based on the success
  of the agent and its neighbors. We measure and describe the
  time evolution of the network and the agents' strategies.
\end{abstract}

\maketitle

\section{Introduction}\label{sec:intro}

To increase or maintain power, or position of influence, 
is a goal of many professionals. Many definitions of power
 recognize that it is not an inherent attribute of an
 actor,\footnote{A person, or other well-defined social unit, in the
   context of our model; we will use the term \textit{agent}.} but a
 result of the interaction between agents. One well-known definition
 by Max Weber reads \cite{weber:power}:

\begin{quote}
`Power' is the probability that one actor within a social relationship
will be in position to carry out his own will despite resistance,
regardless of the basis on which this probability rests.
\end{quote}

Definitions like this suggest that there is a link between the power
of an actor and its position in the network of social relationships.
Thus, by examining a social network, one should be
able to say something about the power of the agents. A major theme in
social network studies has been to infer the power structures in
organizations based on the contact patterns of their
members~\cite{knoke:political}. In undirected networks of actors,
coupled pairwise by their social ties, one idea of measuring, or
defining power, is to say that an actor that is close to others has
more power, than a more peripheral actor does~\cite{sab:clo}. This can be
turned into a network measure called \emph{closeness centrality} (which will
be defined explicitly in the next section). Naively, a way to achieve
power would then be to position oneself as close to everyone else
in the network as possible, i.e.\ to have a social tie to each one of
the network's actors. In practice, to make, and maintain, a social tie
requires the actor to invest time and other resources. To have a
direct tie to a significant fraction of the network is thus neither
feasible, nor desirable. We call this situation of two contrasting
interests---to maximize power (in terms of being central), while at
the same time keeping the number of social ties to a minimum---the
\emph{diplomat's dilemma}.

The diplomat's dilemma can also be motivated from a more academic
perspective. Fueled by the increased availability of large-scale
network datasets, there is a wave of interest in analyzing and
modeling systems as graphs. One theme within this field of
\emph{complex-network theory}~\cite{mejn:rev,ba:rev,doromen:book} has
been to study systems where the network is formed by strategic
decisions by the agents (i.e., situations where the success of the
agents depend on the choice of other agents). This problem has been
traditionally been analyzed from a \emph{game theory} perspective. Some of the most
interesting game-theoretical problems have been inspired by situations where the
agents have conflicting objectives. In, for example, the iterated
prisoner's dilemma~\cite{axe:evo}, agents have to choose between
trying to achieve short-time benefits by exploiting other agents, and
trying to optimize their long-term profit by building a relationship
of mutual trust, but at the same time making them vulnerable to
exploitation. In most real complex networks, and network models, there
is a strong positive correlation between different centrality
measures~\cite{centr:keiko} such as the local degree centrality (the
number of neighbors of a vertex), and closeness centrality. However, one must note that
the correlation between these quantities, though mathematically possible
---high centrality and low degree (and vice versa)--- is not strictly necessary. 
A potentially interesting question in the interface
between complex networks and game theory would then be "How can agents
simultaneously maximize their centrality and minimize their degree?".
Another interesting aspect of this problem, in a more
model-theoretic sense, is that the success of the agents can be
estimated from their network positions alone. In most models of
adaptive, coevolutionary networks~\cite{gross:rev}, the score of
the agents is related to some additional traits of the agents
themselves and their interaction. Our model differs from this approach in the
sense that the success of agents can be measured from the topological features of 
the graph itself, rather than some \emph{extremal} attribute artificially ascribed to the agents.

In this chapter, we will discuss how this problem, the diplomat's
dilemma, can be phrased in more mathematical terms. We will analyze a model of
adaptive agents that try to solve this problem as the network
evolves~\cite{our:hiclod}. We will also, discuss the output of this
model, both the evolution of the network and the evolution of
strategies of the agents.

\section{Definition of the Model}\label{sec:def}

\subsection{Preliminaries}\label{sec:preli}

The framework of our study is a graph $G(t)=\{V(t),E(t)\}$ of $N$
vertices $V$ and $M(t)$ edges $E(t)$.  The vertex set $V$ is fixed, but
the edge set $E(t)$ varies (both its content and size) with time. A
vertex marks the position of an agent in a social network of edges
representing social ties. We will henceforth also assume the graph to
be simple, i.e.\ no multiple- or self-edges are allowed. Let $d(i,j)$
denote the \emph{distance} between $i$ and $j$. Technically we define
$d(i,j)$ as the smallest number of edges in any path (sequence of
adjacent edges) connecting $i$ and $j$. Then, for a connected graph $G$, the
closeness centrality~\cite{sab:clo} is defined as:
\begin{equation}
  c_C(i)=\frac{N-1}{\sum_{j\in G\setminus\{i\}} d(i,j)}.
\end{equation}
The score function, that the agents seek to optimize, should increase
with closeness centrality and decrease with degree. A simple choice for such a
function is $c_C(i)/k_i$ (where $k_i$ is the degree of
$i$). However, we do not want to restrict ourselves to connected networks. If
the network is disconnected, we make the assumption that being a part
of a large component should contribute to a larger centrality. One way
of modifying closeness centrality to incorporate both these aspects
(short distances and being a part of a large
component implies centrality), is to define the centrality $c(i)$ as
\begin{equation}\label{eq:cent}
  c(i)=\sum_{j\in H(i)\setminus\{i\}} \frac{1}{d(i,j)},
\end{equation}
where $H(i)$ is the connected subgraph $i$ belongs to and $d(i,j)$ is
the graph distance between $i$ and $j$. The number of elements in the
sum of Eq.~(\ref{eq:cent}) is proportional to the number of vertices
of $i$'s connected component which gives a positive contribution from large
components. To obtain this property, we use the average reciprocal
distance, rather than the reciprocal average distance (as in the
original definition of closeness centrality). This adjusted definition gives a
higher weight on the count of closer vertices, but captures similar
features as closeness does. 

With the definitions established above, we are now ready to state the score function:
\begin{equation}\label{eq:score}
  s(i) = \left\{\begin{array}{ll}c(i)/k_i & \mbox{if
        $k_i>0$}\\ 0 & \mbox{if $k_i=0$}\end{array}\right. .
\end{equation}

For the purpose of our simulations, the networks we consider will have a initial configuration similar to
Erd\H{o}s-R\'{e}nyi networks~\cite{er:on} with $M_0$ number of edges. In other words,
the network is generated by adding $M_0$ edges one-by-one to $N$ (isolated) vertices
such that no multiple- or self-edge is formed.

\subsection{Moves}

We have outlined so far the basic setup for the game---the underlying graph
representing the actors and their social network, and the score
function that the agents want to optimize. However, to go from this point
to a sensible simulation scheme, we need to determine how an agent can update
its connections. A first, very common assumption, is that the agents are
\emph{myopic}---that they can receive information from, and affect others in
the network only within a certain radius from itself. This assumption lies
behind so much of social network studies that one may argue
that in situations where the myopic assumption is not needed, so
agents can see, and manipulate the network at large distances, the
representation of the social network as a simple graph is not 
appropriate. In our case, we assume that an agent $i$ can change its
connections (affect the network) within the second neighborhood
$\Gamma_2=\{j\in V:d(i,j)\leq 2\}$, and that $i$ can see the
score $s(j)$, centrality $c(j)$ and degree $k_j$ of vertices in $\Gamma_2$. (Since $s$ and $c$ are global quantities, some global
information reach $i$ indirectly. Nevertheless, since the actual contact network
cannot be inferred from this information, we still consider the agents
myopic.)

\begin{figure}
\includegraphics[width=0.8\linewidth]{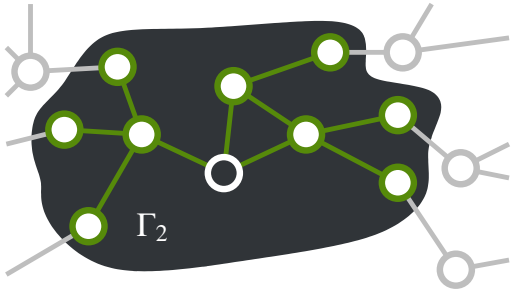}
\caption{An illustration of the myopia (the restricted knowledge about
  the network). The agents are assumed to have knowledge of, and be
  able to affect the second neighborhood $\Gamma_2$ (shaded in the
  figure). The agent knows the centrality and degree of the neighbors
  and their accumulated score the last $t_\mathrm{strat}$ time
  steps. Based on this information the agents can, during a time step,
  based on their strategies, decide to delete the edge to a neighbor,
  and reconnect to a vertex two steps away.
}
\label{fig:ill1}
\end{figure}

The simulations proceed iteratively where, each time step, every vertex
can update its network position by adding an edge to a vertex in
$\Gamma_2$ and delete an edge to a neighbor. An illustration of the
possible moves can be found in Fig.~\ref{fig:ill1}.

\subsection{Strategies}

Ideally one would provide the agents with some
intelligence and use no further restrictions for how they update their
positions to increase their scores.  This is not as easy one can imagine and, for
simplification, one would like to reduce the capability of
the agents further. To do this, we assume that an agent $i$ updates its
position (either by deleting or attaching an edge), by applying a
sequence of tie-breaking \textit{actions}.
\begin{itemize}
\item\textbf{MAXD} Choose vertices with maximal degree.
\item\textbf{MIND} Choose vertices with minimal degree.
\item\textbf{MAXC} Choose vertices with maximal
  centrality in the sense of Eq.~(\ref{eq:cent}).
\item\textbf{MINC} Choose vertices with minimal
  centrality.
\item\textbf{RND} Pick a vertex at random.
\item\textbf{NO} Do not add (or remove) any edge.
\end{itemize}
The sequences of actions define the \emph{strategies} of the
agents. The strategy of an agent $i$ can be stored in two six-tuples
$\mathbf{s}_\mathrm{add}=(s^\mathrm{add}_1,\cdots,s^\mathrm{add}_6)$
and $\mathbf{s}_\mathrm{del}$ representing a priority ordering of the
addition and deletion actions respectively. If $\mathbf{s}_\mathrm{add}(i)=
(\mathrm{MAXD},\mathrm{MINC}, \mathrm{NO},\mathrm{RND},
\mathrm{MIND},\mathrm{MAXC})$ then $i$ tries at first to attach an
edge to the vertex in $\Gamma_2(i)$ with highest degree. If more than
one vertex has the highest degree, then one of these is selected by
the MINC strategy. If still no unique vertex is found, nothing is
done (by application of the NO strategy). Note that such a vertex
is always found after strategies NO or RND are applied. If
$X=\varnothing$ no edge is added (or deleted).

\begin{figure}
\includegraphics[width=\linewidth]{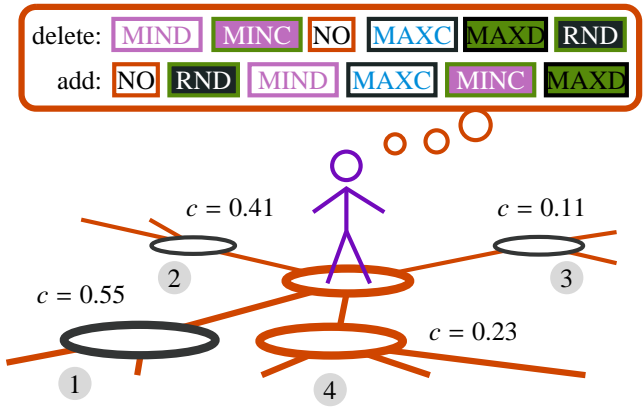}
\caption{An illustration of the strategies of the agents. At a time
  step, the agent can delete one edge and add another in order to
  improve its score. The way to select a neighbor to delete an edge to
  (or a next-nearest neighbor to attach an edge to) is to consecutively
  omit possibilities by applying ``actions'' in a ``strategy
  vector''. This agent's leading deletion strategy is MIND, meaning it
 looks for neighbors with as low degree as
  possible in the first place, to delete the edge to. In this example there are three
  neighbors with degree three (marked with black). To further
  eliminate neighbors the agent applies the MINC strategy (ranking the
  neighbors in order of minimum centrality $c$. In this case
  vertex $1$ is the unanimously least central neighbor. So, at this
  time step, the agent will delete the edge to $1$. As for addition of
  edges the leading action is NO, meaning no edge will be added.
}
\label{fig:ill2}
\end{figure}

\subsection{Strategy updates and stochastic rewiring}

The strategy vectors are initialized to random permutations of the six
actions. Every $t_\mathrm{strat}$'th time step an agent $i$ updates
its strategy vectors by finding the vertex in
$\Gamma_i=\{j:d(i,j)\leq 1\}$ with highest accumulated score
since the last strategy update. This practice of letting the agent mimic
the best-performing neighbor is common in spatial
games~\cite{nowakmay}, and is closely related to the bounded
rationality paradigm of economics~\cite{kah:boundrat}.
When updating the strategy, $i$ copies the parts of
$\mathbf{s}_\mathrm{add}(j)$ and $\mathbf{s}_\mathrm{del}(j)$ that $j$
used the last time step, and let the remaining actions come in the
same order as the strategy vectors prior to the update.
For the purposes of making the set of strategy vectors
ergodic, driving the strategy
optimization~\cite{nowak:wsls,lindgrennordahl}, and modeling
irrational moves by the agents~\cite{kah:boundrat}; we swap,
with probability $p_s$, two random elements of
$\mathbf{s}_\mathrm{add}(j)$ and $\mathbf{s}_\mathrm{del}(j)$ every
strategy vector update.
In addition to the  strategy space we also would like to impose ergodicity in the network space
(i.e.\ the game can generate all $N$-vertex graphs from any initial
configuration). In order to ensure this, disconnected clusters should
should have the ability to reconnect to the graph. We allow this by letting a vertex $i$
attach to any random vertex of $V$ with probability $p_r$ every
$t_\mathrm{rnd}$'th time step. This is not unreasonable as even in real social systems, edges may
form between agents out of sight from each other in the social
network. In fact some authors have pointed out, that in addition to information spreading processes, there are other factors that lead to the evolution of the social networks (cf.\
Ref.~\cite{wattsstrogatz}).

\subsection{The entire algorithm}

To summarize, the algorithm works as follows:
\begin{enumerate}
\item\label{step:init_nwk} Initialize the network to a
  Erd\H{o}s-R\'{e}nyi network with $N$ vertices and $M_0$ edges.
\item\label{step:init_s} For all agents, start with random
  permutations of the six actions as strategy vectors
  $\mathbf{s}_{\mathrm{add}}$ and $\mathbf{s}_{\mathrm{del}}$.
\item\label{step:score} Calculate the score for all agents.
\item\label{step:rewi} Update the agents synchronously by adding
  and deleting edges as selected by the strategy vectors. With
  probability $p_r$, add an edge to a random vertex instead of a
  neighbor's neighbor.
\item\label{step:strat} Every $t_{\mathrm{strat}}$'th time step, update
  the strategy vectors. For each agent, with probability $p_s$, swap
  two elements in it's strategy vector.
\item\label{step:iter} Increment the simulation time $t$. If
  $t<t_\mathrm{tot}$, go to step~\ref{step:score}.
\end{enumerate}
The parameter $n_\mathrm{avg}$, averages over different realizations of the algorithm
are performed. We will primarily use the parameter values $M_0=3N/2$,
$p_s=0.005$, $t_\mathrm{strat}=10$, $t_\mathrm{tot}=10^5$ and
$n_\mathrm{avg}=100$.

\section{Numerical results}\label{sec:res}

\begin{figure}[h]
\includegraphics[width=\linewidth]{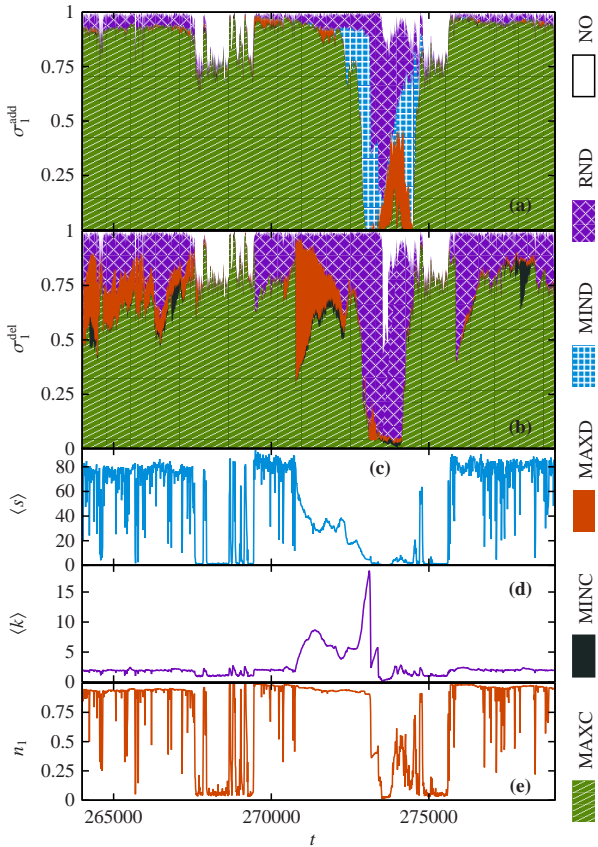}
\caption{
  Output from an example run of a $N=200$ system with $p_r=0.012$. (a)
  and (b) show the fraction of vertices having a certain leading
  action for addition $\sigma_1^{\mathrm{add}}$ (a) and deletion
  $\sigma_1^{\mathrm{add}}$ (b) respectively. (c) shows the average
  score $\langle s\rangle$, (d) the average degree $k$ and (e) the
  fraction of vertices in the largest connected component $n_1$.}
\label{fig:evo}
\end{figure}

\subsection{Time evolution}

To get a feeling for the time evolution, we start by plotting quantities
characterizing the strategies of the agents and the network
structure. The most important parts of the strategy vectors are the
first positions $s^{\mathrm{add}}_1$ and $s^{\mathrm{del}}_1$. In
practice, $\sim 90\%$ of the decisions whether or not to add (or
delete) a specific edge do not pass this first tiebreaker. In
Fig.~\ref{fig:evo}(a) and (b) we can see how complex the
time-evolution of $s^{\mathrm{add}}_1$ and $s^{\mathrm{del}}_1$ can
be. Each sector of the plot corresponds to a leading addition (or
deletion) action, and they have a size in the y-direction proportional
to the fraction of vertices having that leading action value. The time
evolution is complex, having sudden cascades of strategy changes
and quasi-stable periods. Cascades in the leading addition action seem
to be accompanied by cascades in the leading deletion action. The
particular time-window shown in Fig.~\ref{fig:evo} was chosen to
highlight such cascades. For the parameter values of
Fig.~\ref{fig:evo}, cascades involving more than $75\%$ of the vertices
happens about once every $10^5$ time steps.

In Fig.~\ref{fig:evo}(c) we measure the average score function
$\langle s\rangle$.  Being a non-zero-sum game, the value of $\langle s\rangle$ can
vary significantly, a fact which can be seen upon examining the figure. 
Most of the time, the system is close to the
observed maximum $\langle s\rangle \approx 80$. One reason for lower
scores can be seen in Fig.~\ref{fig:evo}(d) where we plot the average
degree $\langle k\rangle$. For some time steps, the network
becomes very dense with an average degree of almost $20$. As high
degree is not desirable, the average score is low during this
period. This rise in degree has, naturally, a corresponding peak in
the leading deletion action NO. Another reason of the occasional dips in
the average score can be seen in  Fig.~\ref{fig:evo}(e) where we
plot the fraction $n_1$ that belongs to the largest connected
component. This quantity is usually close to one, meaning that all
agents are connected (directly or indirectly), but sometimes this fraction
becomes very low. It is harder (than for the high-degree peaks) to see
the corresponding strategic cause for these fragmented states. There
are usually peaks corresponding to NO as the leading addition action, but these are also
accompanied by peaks corresponding to NO as the leading deletion action. As we will see, this
feature becomes less pronounced as the system size increases.

\begin{figure}[h]
\includegraphics[width=\linewidth]{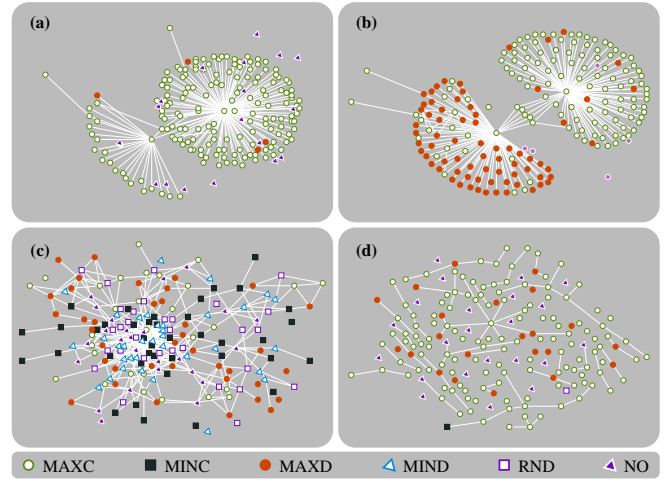}
\caption{Four different example networks from a run with the same
  parameter values as in Fig.~\ref{fig:evo}. The symbols indicate the
  leading addition action. (a) shows the common situation where MAXC is the
  leading addition action. $\sigma_1^{\mathrm{del}}$ is MAXC for
  almost all agents. (b) shows a transition stage between
  $\sigma_1^{\mathrm{add}}$ being mostly MAXD to
  $\sigma_1^{\mathrm{add}}$ being primarily MAXC. (c) shows another
  transient configuration where a large number of different addition
  strategies coexist. (d) shows the addition strategies in a
  fragmented state.
}
\label{fig:nwk}
\end{figure}

\subsection{Example networks}

In light of the complex time evolution of the system, it is not
surprising that the system attains a great variety of network
topologies as time progresses. In Fig.~\ref{fig:nwk} we show four
snapshots of the system for a run with the same parameter values as in
Fig.~\ref{fig:evo}. In Fig.~\ref{fig:nwk}(a) the network comes from
the most common strategy configuration where both the leading deletion
and addition actions are MAXC for a majority of the agents (in this
situation, we call the actions \emph{dominating}). In
this configuration the network is centered around two indirectly
connected hubs. The vertices between these two hubs have the highest
centrality, and since they are within the second neighborhood of most
vertices in the network, and most agents have
$\sigma_1^{\mathrm{add}}=\mathrm{MAXC}$, these vertices will get an
edge from the majority of agents (thus becoming hubs in the next
time-step). There are $18$ isolates with
$\sigma_1^{\mathrm{add}}=\mathrm{NO}$. These will stay isolates until
their strategy vectors are mutated, which occurs (on average) every
$t_{\mathrm{strat}}/p_s=2000$'th time step. Fig.~\ref{fig:nwk}(b)
shows a rather similar network topology with the difference that a
majority of the vertices have MAXD as their leading addition
action (almost all vertices have
$\sigma_1^{\mathrm{del}}=\mathrm{MAXC}$). For this configuration, the
MAXC vertices will move their edges to the most central vertices
whereas the MAXD vertices will not move their edge. In
Fig.~\ref{fig:nwk}(c) we show a more rare, high-$\langle k\rangle$
configuration ($t\approx 273,545$ in Fig.~\ref{fig:evo}). Here the
leading deletion action is NO for about one fourth of the vertices,
and the system is rapidly accumulating edges.  In
Fig.~\ref{fig:evo}(d) we show a fragmented state, where a number of
vertices have the leading addition action NO. The vertices with
$\sigma_1^{\mathrm{add}}=\mathrm{NO}$ that are not isolates have
$\sigma_1^{\mathrm{del}}=\mathrm{NO}$ so they will not fragment the
network further. On the other hand, the vertices with
$\sigma_1^{\mathrm{add,del}}=\mathrm{MAXC}$ and
$\sigma_1^{\mathrm{add,del}}=\mathrm{MAXD}$ \emph{can} fragment the network.

\begin{figure}[b]
\includegraphics[width=\linewidth]{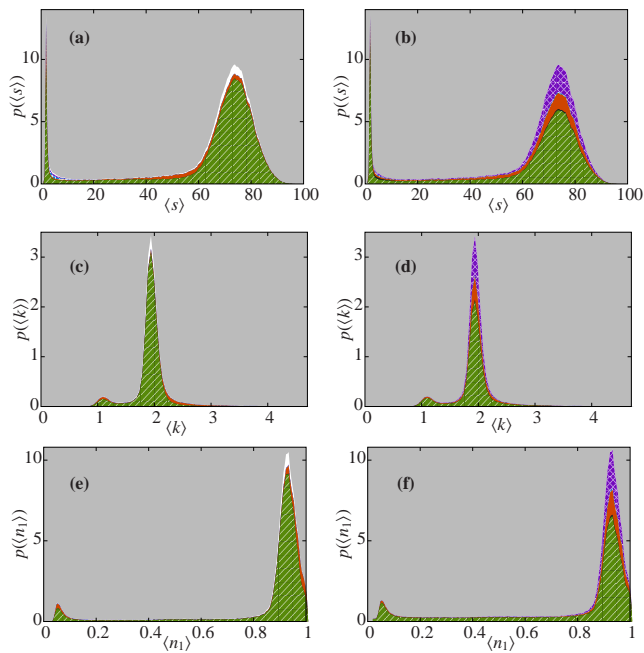}
\caption{ The probability density function of average scores (a),
  (b), average degrees (c), (d), and relative sizes of the largest
  connected component (e), (f). The different fields represent
  different leading addition actions (a), (c), (e), and different
  leading deletion actions (b), (d), (f). The vertical size of a field
  gives the probability density function conditioned to that leading
  action. The curves are averages over ten runs of $10^5$ timesteps
  with the same parameter values as in Fig.~\ref{fig:evo}. The color
  codes of the actions are the same as in Fig.~\ref{fig:evo}.
}
\label{fig:hst}
\end{figure}

\begin{figure}[b]
\includegraphics[width=\linewidth]{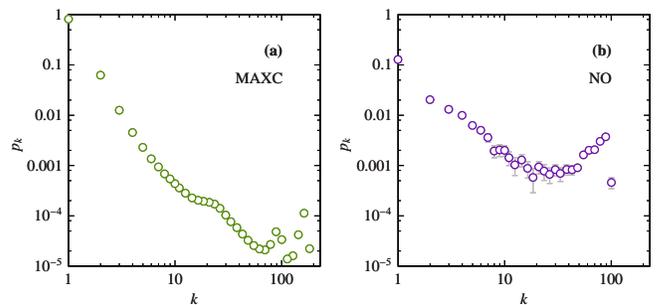}
\caption{
  The degree distribution for systems with the same parameter values
  as in Fig.~\ref{fig:evo}. Panel (a) shows the averaged degree
  distribution when more than half of the agents have MAXC as their
  leading addition actions. Panel (b) displays the corresponding plot for
  the leading addition action NO.
}
\label{fig:deg}
\end{figure}

\subsection{Effects of strategies on the network topology}

We are now in a position to examine in detail the network
topologies that arise from different dominating addition and deletion
actions. First, we plot histograms (rescaled to show the probability
density functions) of the network structural quantities shown in
Figs.~\ref{fig:evo}(c), (d) and (e)---see Fig.~\ref{fig:hst}. These diagrams all have
two peaks---one with low $\langle s\rangle$,  $\langle k\rangle$ and
$\langle n_1\rangle$ values (where the network is fragmented, the
number of edges small and the scores low), and another broader peak
corresponding to a connected network with higher scores and more
edges. Interestingly, the different leading actions are not completely
localized to different peaks but spread out over the whole
range. Another counter-intuitive observation is that there seems to be
more agents with $\sigma_1^{\mathrm{add}} = \mathrm{NO}$ in the more
dense peaks. These vertices (with $\sigma_1^{\mathrm{add}} =
\mathrm{NO}$) seem to be primarily isolated and do not affect the
majority of vertices (connected in the largest component). They will
therefore stay isolated until their strategies have changed or they
have been connected to the rest of the network by random
connections. We also observe that there is a larger variety of
leading addition actions than leading deletion actions. A possible interpretation of this
is that the fitness of agents is more dependent on the leading addition action. This seems natural
in a situation where it is disadvantageous to connect to a majority of agents
(so the choice of neighbor to disconnect is not important), however it is beneficial to connect to 
to a minority of well established agents.

One of the most widely studied and revealing metrics of network structure, is the degree
distribution---the probability mass function of the degrees of
vertices. In Fig.~\ref{fig:deg} we plot the degree distribution for
dominating actions
$\sigma_1^{\mathrm{add}} = \mathrm{MAXC}$ (a) and
$\sigma_1^{\mathrm{add}} = \mathrm{NO}$ (b). The
$\sigma_1^{\mathrm{add}} = \mathrm{MAXC}$ graph has two high-$k$
peaks, corresponding to the hubs in the network. The existence of two broad peaks as opposed to only one is strange, and the reasons for this is not immediately apparent. The
$\sigma_1^{\mathrm{add}} = \mathrm{NO}$ graphs (whose averaged degree
distribution are shown in (b)) are more dense, as expected. However, they
also have a large-$k$ peak, which is probably related to, either the
strategies of other agents, or a residue from the preceding period
(remember that the periods of dominating $\sigma_1^{\mathrm{add}} =
\mathrm{NO}$ is very short compared with the $\sigma_1^{\mathrm{add}}
= \mathrm{MAXC}$ periods). This implies that one can separate
system-wide effects of some strategy driving the decisions of the
majority, but there will also be other effects present in the network.
Note that, while many studies have focused on the emergent properties of
degree distributions as $N\rightarrow\infty$, the interesting features of our
model occurs for smaller system sizes, consequently we believe this limit is not
interesting or relevant to our study and we do not consider it.

\begin{figure}[b]
\includegraphics[width=\linewidth]{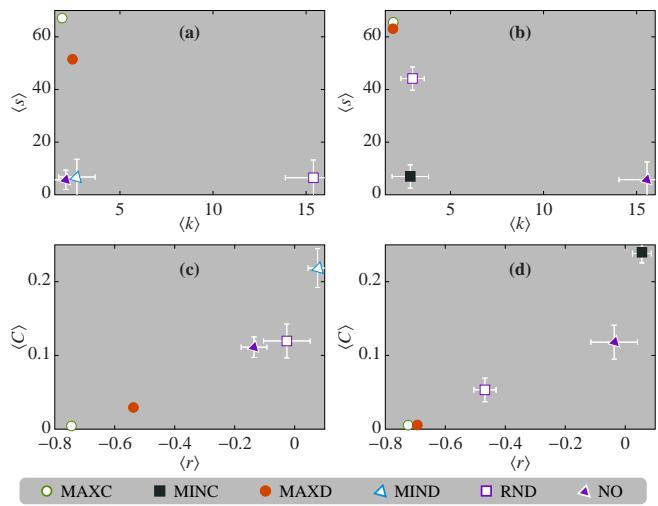}
\caption{
  Average values of four different network structural quantities for
  different dominating addition and deletion actions (i.e.\ that more
  than half of the agents have a specific $\sigma_1^{\mathrm{add}}$,
  or $\sigma_1^{\mathrm{del}}$). (a) shows the average score as a
  function of degree for different dominating
  $\sigma_1^{\mathrm{add}}$. (b) is the corresponding plot for
  $\sigma_1^{\mathrm{del}}$. (c) displays the clustering coefficient
  as a function of assortativity for different dominating
  $\sigma_1^{\mathrm{add}}$. (d) is the corresponding plot for
  different $\sigma_1^{\mathrm{del}}$. The bars indicate standard
  errors. The data comes from simulation of ten runs (different random
  number generator seeds) of $10^5$ time steps. Two actions were never
  attained during these runs: $\sigma_1^{\mathrm{add}}=\mathrm{MINC}$
  and  $\sigma_1^{\mathrm{del}}=\mathrm{MIND}$.
}
\label{fig:nws}
\end{figure}

We now proceed to look at four other measures of different network
structures and how they depend on the dominating addition and deletion
actions. The first two measures we consider are the degree $k$ and score $s$. In
Fig.~\ref{fig:nws}(a) and (b) we plot the average values of these
quantities (averaged over all vertices, regardless of strategy, and
averaged over all samples with a particular dominating strategy). This
plot is based on ten runs for $10^5$ time steps, with network
quantities measured every tenth time step. During these runs, the two leading actions---$\sigma_1^{\mathrm{add}}=\mathrm{MINC}$
and $\sigma_1^{\mathrm{del}}=\mathrm{MIND}$ were never employed. 
We note that the most common leading actions (for both addition and deletion) MAXC and MAXD
gives the highest average score. This does not mean that all agents
have a high score in these situations---from
Figs.~\ref{fig:nwk}(a) and (b) we know that the score can differ much
from one agent to another. The degrees are low for these strategies,
which is a necessary (but not sufficient) condition for a low
score. For $\sigma_1^{\mathrm{add}}=\mathrm{NO}$ the average degree is
also low, but the score is much lower than for
$\sigma_1^{\mathrm{add}}=\mathrm{MAXC}$ and MAXD. The reason, as
pointed out above, is that the network can become heavily fragmented
for this leading addition action. The $\sigma_1^{\mathrm{add}}$ corresponding
to the highest degree is RND, this might seem strange, but during these
runs (which is also visible in Fig.~\ref{fig:evo})
$\sigma_1^{\mathrm{add}}=\mathrm{RND}$ is correlated with
$\sigma_1^{\mathrm{del}}=\mathrm{NO}$ which is a state naturally
leading to a comparatively dense network. The other leading actions
$\sigma_1^{\mathrm{add}}=\mathrm{MIND}$ and
$\sigma_1^{\mathrm{del}}=\mathrm{MINC}$ result in low scores and sparse
networks.

The other two measures we examine are the
\textit{assortativity} and \textit{clustering coefficient}. Before
discussing our results, let us first define these quantities in detail. 
The average degree tells us if the network is sparse or dense. The
degree distribution gives a more nuanced picture of how
homogeneous the set of vertices are with respect to the number of
neighbors. The next level of complexity in describing the network with respect to the agents'
degree, is to measure the correlations between the degrees of vertices
at either side of an edge. In particular, one can determine if high-degree vertices are
primarily connected to similar high degree vertices, or instead are
linked to low-degree vertices.  The assortativity $r$ is a measure of vertices' tendency to connect to other vertices of similar type, in this case those with similar degree~\cite{mejn:rev}.  In technical terms, $r$ is the Pearson correlation coefficient of the degrees at either side of
an edge. There is an additional caveat that we need to consider; since the edges in our networks are
undirected, $r$ has to be symmetric with respect to edge-reversal
(i.e.\ replacing $(i,j)$ by $(j,i)$).  However the the standard definition of the Pearson correlation coefficient does not account for this symmetry.
The way to fix this problem is to let one edge contribute twice to
$r$, i.e.\ to represent an undirected edge by two directed
edges pointing in opposite directions. If one employs an edge list
representation internally (i.e., if edges are stored in an array
of ordered pairs $(i_1,j_1),\cdots,(i_M,j_M)$) then we can write the adjusted $r$ as,
\begin{equation}\label{eq:asso}
  r=\frac{4\langle k_1\, k_2\rangle - \langle k_1 + k_2\rangle^2}
  {2\langle k_1^2+k_2^2\rangle - \langle k_1+ k_2\rangle^2},
\end{equation}
where, for a given edge $(i,j)$, $k_1$ is the degree of the first argument
(i.e., the degree of $i$), $k_2$ is the degree of the second
argument and the brackets $\langle\cdots\rangle$ denote averaging. The range of $r$ is $[-1,1]$ where negative values indicate a preference for highly connected vertices to attach to low-degree
vertices, and positive values imply that vertices tend to be attached
to other vertices with degrees of similar magnitudes.

The clustering coefficient, on the other hand, is a measure of transitivity in the network.  In other words it checks whether neighbors of a node are also connected to each other (thus forming triangles).  
It is a well known empirical fact that social acquaintance networks have a
strong tendency to form triangles~\cite{holl:72} and it is therefore a
worthwhile exercise to examine whether the networks generated by our
model display this feature. There is in principle, more than one way
to define the clustering coefficient.  Here we employ the most
commonly used one~\cite{bw:sw},
\begin{equation}\label{eq:clust}
  C = 3 n_\mathrm{triangle}\:\big/\:n_\mathrm{triple},
\end{equation}
where $n_\mathrm{triangle}$ is the number of triangles and
$n_\mathrm{triple}$ is the number of connected triples (subgraphs
consisting of three vertices and two or three edges). The factor of three
is included to normalize the quantity to the interval $[0,1]$.

Now that we have defined these quantities we refer back to Fig.~\ref{fig:nws}.  
We note that the most common leading
actions $\sigma_1^{\mathrm{add,del}}=\mathrm{MAXC}$ and MAXD have the
lowest $\langle C\rangle$ and $\langle r\rangle$ values.  A possible explanation for this could be the following. Consider a triangle, a subgraph of three vertices connected by three edges.  The
graph will be connected even if one of these edges is deleted. In a
situation where edges are expensive, this kind of redundancy is not
desired. For this reason, it seems natural that, on average, the
most successful strategies MAXC and MAXD have few triangles. The
negative assortativity of these situations are also conspicuous features of the
examples shown in Figs.~\ref{fig:nwk}(a) and (b) (most vertices there are only
connected to the two hubs, but the hubs are not connected to each
other). For networks with a broad spectrum of degrees, it is known
that $\langle C\rangle$ and $\langle r\rangle$ are relatively strongly
correlated~\cite{hz:ac}. This is also true in Figs.~\ref{fig:nws}(c)
and (d) where the relationship between $\langle C\rangle$ and
$\langle r\rangle$ is monotonically increasing. The network configurations
with highest $\langle C\rangle$ and $\langle r\rangle$ are the ones
with $\sigma_1^{\mathrm{add}}=\mathrm{MIND}$ and
$\sigma_1^{\mathrm{del}}=\mathrm{MINC}$. Since these networks are both
sparse and  fragmented, some components must have a large number of triangles 
(probably close to being fully connected). The denser states, with
$\sigma_1^{\mathrm{add}}=\mathrm{RND}$ and
$\sigma_1^{\mathrm{del}}=\mathrm{NO}$, have intermediate $\langle
C\rangle$- and $\langle r\rangle$-values, meaning that the edges are more
homogeneously spread out, similar to the network in
Fig.~\ref{fig:nwk}(c).

\subsection{Transition probabilities}

From Fig.~\ref{fig:evo} it seems likely that the ability of one leading
action to grow in the population depends on the other predominant
strategies in the system. For example, $\sigma_1^{\mathrm{add}}=\mathrm{RND}$
dominates after a period of many agents employing
$\sigma_1^{\mathrm{add}}=\mathrm{MIND}$ as the leading strategy. Consequently, it is worth asking the question: How does the probability of one
leading action depend on the configuration at earlier time steps?

\begin{table*}
\begin{ruledtabular}
\begin{tabular}{r|cccccc}
& MAXC & MINC & MAXD & MIND & RND & NO \\\hline
MAXC & 1 & 0.0164(3) & 0.0088(2) & 0.0107(4) & 0.0151(5) & 0.0010(0)\\
MINC & 0.0169(3) & 1 & 0.0113(6) & 0.036(2) & 0.025(2) & 0.0017(3)\\
MAXD & 0.0093(3) & 0.0104(7) & 1 & 0.0103(6) & 0.0206(9) & 0.0003(0)\\
MIND & 0.0115(4) & 0.030(2) & 0.0130(7) & 1 & 0.059(5) & 0.0020(2)\\
RND & 0.0157(5) & 0.024(2) & 0.020(1) & 0.064(5) & 1 & 0.0023(5)\\
NO & 0.0007(0) & 0.0031(2) & 0.0009(0) & 0.0036(2) & 0.0042(4) & 1\\
\end{tabular}
\end{ruledtabular}
\caption{Values for the $\mathbf{T}$ matrices for addition.
  ($T_{ij}$ is the deviation from the expected value in
  a model of random transitions given the diagonal values.)
  The values are averaged over 100 realizations of the
  algorithm. All digits are significant to one s.d. The parameter
  values are the same as in Fig.~\ref{fig:evo}. Numbers in parentheses
  are the standard errors in units of the last decimal.}
\label{tab:add}
\end{table*}

\begin{table*}
\begin{ruledtabular}
\begin{tabular}{r|cccccc}
  & MAXC & MINC & MAXD & MIND & RND & NO\\\hline
  MAXC & 1 & 0.0100(2) & 0.0131(4) & 0.0094(2) & 0.0266(3) & 0.0126(3)\\
  MINC & 0.0098(2) & 1 & 0.0070(3) & 0.010(1) & 0.0105(4) & 0.0050(3) \\
  MAXD & 0.0133(4) & 0.0067(3) & 1 & 0.0055(2) & 0.0124(3) & 0.0062(2)\\
  MIND &0.0087(2) & 0.011(1) & 0.0054(2) & 1 & 0.0101(2) & 0.0055(3)\\
  RND &0.0269(3) & 0.0094(4) & 0.0128(3) & 0.0083(2) & 1 & 0.0072(3)\\
  NO &0.0097(3) & 0.0076(3) & 0.0053(2) & 0.0078(3) & 0.0131(3) & 1\\
\end{tabular}
\end{ruledtabular}
\caption{Same as in Tab.~\ref{tab:add} but for deletion, instead of
  addition, strategies. }
\label{tab:del}
\end{table*}

We investigate this qualitatively by calculating the ``transition matrix'' $\mathbf{T}'$
with elements $T'(s_1,s_1')$ giving the probability of a vertex with
the leading action $s_1$ to have the leading action $s_1'$ at the
next time step. However, note that the dynamics is not fully determined
by $\mathbf{T}'$, and is thus not a transition matrix in the sense of
other physical models. If that were the case (i.e.\ the current strategy is
independent of the strategy adopted in the previous time step) we would have
the relation $T'_{ij}=\sqrt{T'_iT'_j}$. To study the deviation from
this null-model, we assume the diagonal (i.e.\ the frequencies of
the strategies) given, and calculate $\mathbf{T}$ defined by,
\begin{equation}\label{eq:theta}
  T_{ij} = T'_{ij} / \sqrt{T'_iT'_j}.
\end{equation}
The values of $\mathbf{T}$ for the parameters defined in
Fig.~\ref{fig:evo} are displayed in Tabs.~\ref{tab:add} and
\ref{tab:del}. The off-diagonal elements have much lower values than $1$ (the
average off-diagonal $\Theta$ values are $0.014$ for addition
strategies and $0.010$ for deletion). This reflects the contiguous
periods of one dominating action. Note that transitions between MAXC
and RND are over-represented: $T'^{\mathrm{del}}_{\mathrm{MAXC},
  \mathrm{RND}}\approx T'^{\mathrm{del}}_{\mathrm{RND},
  \mathrm{MAXC}}\approx 0.027$, which is more than twice the value of
any other off-diagonal element involving MAXC or RND. As another token
of the problem's complexity, the matrix is not completely symmetric
$T'^{\mathrm{del}}_{\mathrm{RND}, \mathrm{NO}}$ is twice ($\sim 3$
s.d.)\ as large as $T'^{\mathrm{del}}_{\mathrm{NO}, \mathrm{RND}}$
meaning that it is easier for RND to invade a population with NO as a
leading deletion action, than vice versa.

\begin{figure}[h]
\includegraphics[width=\linewidth]{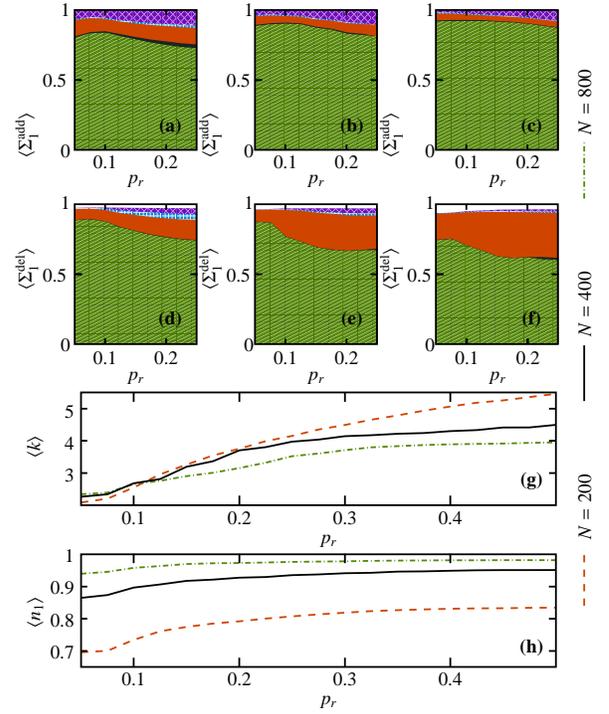}
\caption{The system's dependence on the topological noise level (via
  the fraction of random rewirings $p_r$) for different system sizes
  $N$. Panels (a), (b) and (c) show the fraction
  $\langle\Sigma_1^{\mathrm{add}} \rangle$ of leading addition actions
  $\sigma_1^{\mathrm{add}}$ for systems of $N=200$, $400$ and
  $800$. Panels (d), (e) and (f) show the fraction of preferred
  deletion actions for the same three system sizes, while (g) shows
  the average degree and (h) the average size of the largest connected
  component $\langle n_1\rangle$.
}
\label{fig:par}
\end{figure}

\subsection{Dependence on system size and noise}

So far we have focused on one set of parameter values. In this section
we investigate how the system behavior depends on the number of
agents and the noise level in the deletion and attachment mechanism. In
Fig.~\ref{fig:par}, we tune the noise level (fraction of random
attachments) $p_r$ for three system sizes. In panels (a)--(c) we show
the fraction of leading addition actions among the agents
$\langle\Sigma_1^{\mathrm{add}}\rangle$  (averaged over $\sim
100$ runs and $10^5$ time steps).  The quantities $\Sigma_1^{\mathrm{add,del}}$
denotes the fraction of agents having a specific
$\sigma_1^{\mathrm{add,del}}$. As observed in Fig.~\ref{fig:evo}(a)
the leading action is MAXC followed by MAXD and RND. The leading
deletion actions, as seen in panels (d)--(f), are ranked similarly
except that MAXD has a larger (and increasing) presence. If $p_r=1$,
then all actions are equally likely (they do not have any meaning---all strategies
will result in random moves equal to $s_1^{\mathrm{add}} = s_1^{\mathrm{del}}
=\mathrm{RND}$). There are
trends in the $p_r$-dependences of $\langle \sigma_1^{\mathrm{add}}
\rangle$, but apparently no emerging discontinuity. This observation,
(which also seems to hold for the $p_s$-scaling), that there is no
phase transition for any parameter value governing the probability
of random permutations in the strategy vectors, is an indication that
the results above can be generalized to a large parameter range. We
also note that, although the system has the opportunity to be passive
(i.e.\ agents having $s_1^{\mathrm{add}} = s_1^{\mathrm{del}}
=\mathrm{NO}$), this does not happen. This situation is reminiscent of
the ``Red Queen hypothesis'' of evolution~\cite{redqueen}---organisms
need to keep evolving to maintain their fitness.

Next we look at the dependence of the network structure on the number
of agents and the noise level. The average degree, plotted in
Fig.~\ref{fig:evo}(g) is monotonously increasing with $p_r$. There is,
however, a qualitative difference in the size scaling---for
$p_r\lesssim 0.12$ the average degree increases with $N$, for
$p_r\gtrsim 0.12$ this situation is reversed. In Fig.~\ref{fig:evo}(h)
we plot the average largest-component size as a function of $p_r$ for
different system sizes. The behavior is monotonous in both $p_r$ and
$N$---larger $p_r$, or a larger system size, means higher $\langle
n_1\rangle$. In all network models we are aware of (allowing
fragmented networks), a decreasing average degree implies a smaller
giant component. For $p_r\gtrsim 0.12$, in our model the picture is
the opposite---as the system grows the giant component spans an
increasing fraction of the network. This also means that the agents,
on average, reach the twin goals of keeping the degree low and the
graph connected.

\section{Discussion}
\label{sec:disc}

We have presented a general game theoretic network problem, the
diplomat's dilemma---how can an agent in a network simultaneously
maximize closeness centrality and minimize degree. The motivation for
this problem comes in part from a type of social optimization situation
where agents seek to gain power (via closeness centrality) and keep
the cost (degree) low. It can also be motivated from a more academic
point of view---interesting dynamics often comes from when agents
simultaneously try to optimize conflicting objectives. The diplomat's
dilemma is one of the simplest such situations in a networked system,
because the score function does not depend on any additional variable,
or trait, of the vertices, only the vertex' position in the network.

We devise an iterative simulation where at every time step, an agent can delete it's connection
to a neighbor and add an edge to a second neighbor, based on
the information it possesses about the network
characteristics of vertices within its local neighborhood (upto second neighbors). The agents use
strategies that they update by imitating the best performing neighbor
within this information horizon. The dynamics
are driven by occasional random moves and random permutations of the
vectors encoding the strategies of the agents. 
For the sake of flexibility, the definition of the problem as stated in this chapter, is deliberately vague. To turn it
into a mathematically well-defined problem, one has to specify how the
agents can affect their position in the network and what information
they can use for this objective. There are  of course many choices for how to do this.
Although we believe our formulation is natural, it would be very
interesting to rephrase these assumptions. 
A future enhancement would be to equip the agents with methods from the machine
learning community to optimize their position, and to tune the amount
of information accessible to the agents. A mathematical simplification
of the problem would be to let all agents know the precise network
topology at all times (this may however lead to some conceptual
problems---if the information about the network is obtained via the
network, it would be strange if the picture of the network close to an
agent would not be more accurate than the picture of more remote
sections of the network). Another interesting version of the problem
would be to require an edge to represent an agreement between both vertices,
so that an agent $i$ cannot add an edge $(i,j)$ unless $j$ finds this
profitable.

Nevertheless, despite the simplicity of our model, the time evolution of the simulation is strikingly complex, with
quasi-stable states, trends, spikes and cascades of strategies among
the agents. This complex dynamics is also captured in various metrics
measuring different levels of network structure. Furthermore, the network structure
and the agents' strategies directly influence one another. If the agents
stop deleting edges, the average degree of the network will grow
rapidly, which may benefit a strategy aiming to lower the degree of the
agents. This feedback from network structure, to the agents and their
decisions about how to update their networks is a central theme in the
field of adaptive, coevolutionary networks~\cite{gross:rev}. We believe
that all forms of social optimization involve such feedback loops,
which is a strong motivation for studying adaptive networks.
The complexity of the time-evolution, especially in the network
structural dynamics is more striking for intermediate system sizes.
Indeed, many interesting features of our simulation are not emergent
in the large-system limit, but rather present only for small
sizes. Models in theoretical physics have traditionally focused on
properties of the system as $N\rightarrow\infty$. In models of social
systems however, extrapolating to infinite size is not necessarily a natural limit in the same way
(it will of course be interesting to examine the limiting behavior of
such models). We believe this is a good example of the dangers of taking
the large-size limit by routine---the most interesting relevant
features of the model may be neglected.

In a majority of the cases in our simulations, most of the agents use a
strategy where they both delete, and attach to vertices according to
the MAXC action. This implies that the agent first deletes the edge to
the most central vertex in the second neighborhood (in the sense of a
modified closeness centrality), and then reattaches to the most central
vertex two steps away (before the deletion). In practice this means
that an agent typically transfers an edge from it's  most central immediate neighbor to
it's most central neighbor two steps away. This strategy makes the agent move
towards the center without increasing its degree, which clearly seems
like a reasonable procedure in the diplomat's dilemma. However, this
strategy is not evolutionary stable in the presence of noise (hence
the complex time evolution). This strategy creates networks with
low clustering coefficients, i.e., there are a comparatively small
number of triangles. Since forming a triangle introduces an extra edge, which
is expensive, without changing the size of connected component, one
can understand why agents are reluctant to form these triangles \emph{per se} in our formulation of the problem.

Different strategies have different ability to invade one another. To
test this we measure the deviation from random transitions from one
dominating action to another (given the frequency of particular
strategies), concluding for example that it is about twice as easy for
RND to invade NO as a leading deletion action. Another interesting
aspect is that (for some noise levels), as the system size increases,
the network becomes both more connected (the relative fraction of
vertices in the largest connected component increases), and more
sparse (the average degree decreases). This is in sharp contrast to all
other generative network models that we are aware of, but definitely consistent with the objectives of the general problem (where large connected components and low
degrees are desired).

What does this result tell us about the real professional life of
diplomats? Maybe that they can, by selfishly optimizing their
positions in the network, self-organize to a connected business
network where they need only a few business contacts, without knowing
more about the network than the second neighborhood. However to make a stronger and more conclusive statement about the optimal strategy, more results are needed. This is
something we hope to gather from future studies. 

One of the problems facing this type of mechanistic modeling of social information
processes~\cite{our:coevo,zanet:coevo,rosv:soc,rosvall1,mama:soc}, is that they are very hard to validate. Information spreading in social systems is neither routed from agent to agent like
the information packets in the Internet, nor do they spread in the same fashion as epidemics. 
Instead the spreading dynamics is content
dependent. Different types of information may be spreading over
different social networks, following different dynamic rules. There
are some promising datasets for studying social information
spreading. For example, networks of blogs, Internet communities, or
social networking sites generate large amounts of potentially valuable data, although
these data sets are not necessarily conducive to the questions that adaptive
models such as the one described in this chapter seek to address. In
the near future, we hope mechanistic modeling of social information
processes will be more data driven, asking questions that can actually
be validated through empirical study.

\acknowledgements{
  P.H. acknowledges support from the Swedish Foundation for Strategic
  Research. G.G. thanks the James S. McDonnell foundation for support.
The authors thank Matteo Marsili for comments.
}

\end{document}